\documentclass[fleqn,usenatbib]{mnras}
\usepackage{graphicx}	% Including figure files
\usepackage{amsmath}	% Advanced maths commands
\usepackage{txfonts}
\usepackage[T1]{fontenc}
\usepackage{ae,aecompl}
\title[Halo Meteors]{Halo Meteors}

\author[A. Siraj and A. Loeb]{
Amir Siraj,$^{1}$\thanks{amir.siraj@cfa.harvard.edu}
Abraham Loeb,$^{1}$\thanks{aloeb@cfa.harvard.edu}
\\
$^{1}$Department of Astronomy, Harvard University, 60 Garden Street, Cambridge, MA 02138, USA\\
}

\date{Accepted XXX. Received YYY; in original form ZZZ}

\pubyear{2019}

\begin{document}
\label{firstpage}
\pagerange{\pageref{firstpage}--\pageref{lastpage}}
\maketitle

% Abstract of the paper
\begin{abstract}
The stellar halo contains some of the oldest stars in the Milky Way galaxy and in the universe. The detections of `Oumuamua, CNEOS 2014-01-08, and interstellar dust serve to calibrate the production rate of interstellar objects. We study the feasibility of a search for interstellar meteors with origins in the stellar halo. We find the mean heliocentric impact speed for halo meteors to be $\sim 270 \mathrm{\;km\;s^{-1}}$, and the standard deviation is $\sim 90 \mathrm{\;km\;s^{-1}}$, making the population kinematically distinct from all other meteors, which are an order-of-magnitude slower. We explore the expected abundance of halo meteors, finding that a network of all-sky cameras covering all land on Earth can take spectra and determine the orbits of a few hundred halo meteors larger than a few mm per year. The compositions of halo meteors would provide information on the characteristics of planetary system formation for the oldest stars. In addition, one could place tight constraints on baryonic dark matter objects of low masses.
\end{abstract}

% Select between one and six entries from the list of approved keywords.
% Don't make up new ones.
\begin{keywords}
Galaxy: halo -- meteorites, meteors, meteoroids -- planetary systems -- minor planets, asteroids: general -- comets: general
\end{keywords}

%%%%%%%%%%%%%%%%%%%%%%%%%%%%%%%%%%%%%%%%%%%%%%%%%%

%%%%%%%%%%%%%%%%% BODY OF PAPER %%%%%%%%%%%%%%%%%%

\section{Introduction}
`Oumuamua was the first interstellar object detected in the Solar System by Pan-STAARS \citep{Meech2017, Micheli2018}. Several follow-up studies of `Oumuamua were conducted to better understand its origin and composition \citep{Bannister2017, Gaidos2017, Jewitt2017, Mamajek2017, Ye2017, Bolin2017, Fitzsimmons2018, Trilling2018, Bialy2018, Hoang2018, Siraj2019a, Siraj2019b, Seligman2019}. 

There is significant evidence for previous detections of interstellar meteors \citep{Baggaley1993, Hajdukova1994, Taylor1996, Baggaley2000, Mathews1998, Meisel2002a, Meisel2002b, Weryk2004, Afanasiev2007, Musci2012, Engelhardt2017, Hajdukova2018}, including the meter-size CNEOS 2014-01-08 meteor \citep{Siraj2019c}. 

Spectroscopy of gaseous debris from interstellar meteors as they burn up in the Earth's atmosphere could reveal their composition, and a worldwide network of all-sky cameras would allow for the detection and analysis of interstellar meteors at the centimeter-scale \citep{Siraj2019d}.

In this \textit{Letter}, we explore the theoretical population of interstellar meteors originating from the stellar halo, which contains the oldest stars in the Milky Way galaxy and  in the universe \citep{Helmi2008, Frebel2007}. Studying the composition of halo star ejecta could help reveal the nature of primordial planetary system formation. We discuss the kinematics of halo meteors, their abundance and expected detection rates, and what we could learn from analyzing their composition. 

\section{Kinematics}
\label{sec:kinematics}
We assume that the velocity distribution of halo interstellar objects follows that of their parent stars. The mean rotation speed of the Local Standard of Rest (LSR) with respect to the halo is $v_{\phi}^{\odot} \approx 238 \mathrm{\;km\;s^{-1}}$ \citep{Reid2016, Bland-Hawthorn2016}. The Sun's velocity components with respect to the LSR, in right-handed Galactic coordinates, are $(v_U^{\odot}, v_V^{\odot}, v_W^{\odot}) = (10\pm1, 11\pm2, 7\pm5) \mathrm{\;km\;s^{-1}}$ \citep{Bland-Hawthorn2016}. The mean rotation of the local halo population with respect to the LSR is $\bar{v}_{\phi}^{h} \approx 40 \mathrm{\;km\;s^{-1}}$, and the spherical velocity ellipsoid for the local halo population is defined by the velocity dispersion, $(\sigma_r^h, \sigma_{\theta}^h, \sigma_{\phi}^h) = (141\pm5, 75\pm5, 85\pm5) \mathrm{\;km\;s^{-1}}$ \citep{Bland-Hawthorn2016}.

\begin{figure}
  \centering
  \includegraphics[width=\linewidth]{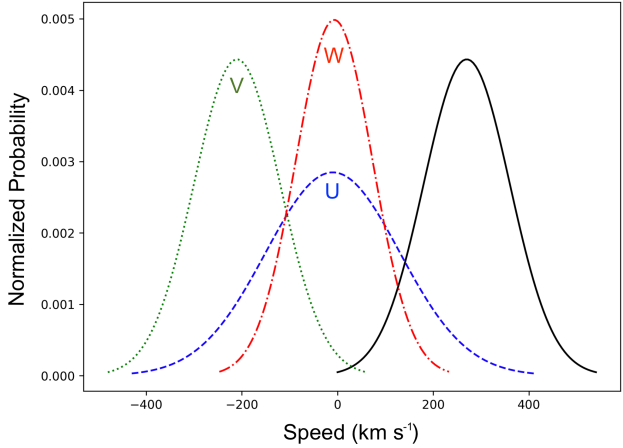}
    \caption{Distribution of heliocentric impact velocities of halo meteors, with U, V, and W, components shown in blue, green, and red, respectively, and total speed displayed in black. Cutoffs are at $\pm 3 \sigma$, where $\sigma$ is the standard deviation.}
    \label{fig:1}
\end{figure}

We use the following Monte Carlo method to determine the heliocentric impact velocities of halo interstellar objects. First, we draw randomly from the Gaussian distributions described by the spherical velocity ellipsoid for the local halo distributions and determine the velocity components of a random halo interstellar object with respect to the LSR. We then subtract the Sun's velocity components relative to the LSR. Finally, assuming an isotropic distribution, we add the kinetic energy from the change in gravitational potential. We find the heliocentric impact velocity distriution to have mean velocities $(\bar{v}_U^h, \bar{v}_V^h, \bar{v}_W^h) = (-10, -210, -7) \mathrm{\;km\;s^{-1}}$, with velocity dispersions $(\sigma_U^h, \sigma_V^h, \sigma_W^h) = (140, 90, 80) \mathrm{\;km\;s^{-1}}$, shown  in Figure~\ref{fig:1}. The mean heliocentric impact speed is $\sim 270 \mathrm{\;km\;s^{-1}}$, and the standard deviation is $\sim 90 \mathrm{\;km\;s^{-1}}$, making halo meteors kinematically distinct from all other (Solar System or interstellar) meteors which have characteristic impact speeds that are an order-of-magnitude smaller.

\section{Abundance \& Detectability}
\label{sec:abundance}
Approximately 1 in every $10^3$ local stars, and therefore interstellar meteors, originate from the Milky Way halo \citep{Helmi2008}. On average, halo meteors travel $\sim5$ times faster than disk meteors. We therefore estimate the impact rates for halo meteors to be a factor of $\sim 200$ lower than those for typical interstellar meteors of the same size \citep{Siraj2019c}.

All-sky camera systems such as AMOS can obtain spectra for typical meteors down to a size scale of $\sim 1$cm, and considering that halo interstellar meteor impact speeds are approximately ten times higher than typical meteors, we estimate that, with an adequate frame rate, a system with the sensitivity of AMOS could obtain spectra for halo meteors down to a size scale of $\sim2$mm \citep{Toth2015}.

\begin{figure}
  \centering
  \includegraphics[width=\linewidth]{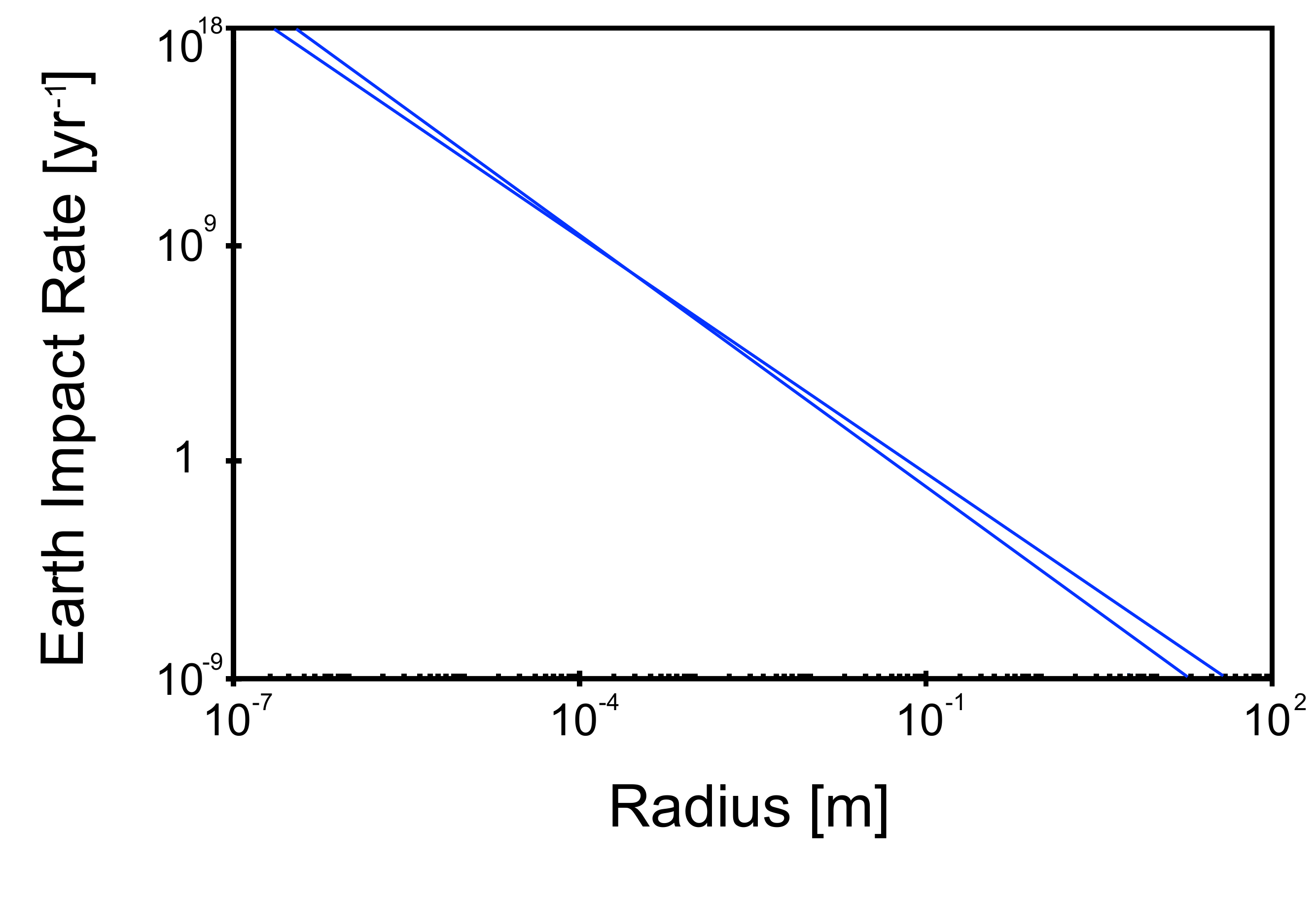}
    \caption{Range of expected earth impact rates of halo meteors as a function of size, based on the $q = 3.41 \pm 0.17 $ estimate \citep{Siraj2019c}. }
    \label{fig:2}
\end{figure}

Figure~\ref{fig:2} shows expected Earth impact rates as a function of size for halo meteors, assuming a similar size distribution for ejected mass as disk stars \citep{Siraj2019c}. We expect 2mm halo meteors to impact the Earth at a rate of $10^4 \mathrm{\;yr^{-1}}$. A network of all-sky cameras covering all land on Earth, with the sensitivity of AMOS and an adequate frame rate to determine the orbits of $\sim 300 \mathrm{\;km\;s^{-1}}$ impactors, are expected to detect and take spectra for nearly $10^3$ halo meteors per year. We would not expect to obtain physical samples of halo meteors, as meteors larger than 10cm are expected to impact the Earth once every few hundred years. Finally, the ratio of disk to halo meteors will inform our understanding of primoridal planetary system formation.

\section{Implications}
\label{sec:implications}
\subsection{Early Planetary Systems}
There are three major phases of interstellar object production during a star's lifetime \citep{Pfalzner2019}. The first is the shedding of icy planetesimals from the outer disk by interactions with neighboring stars \citep{Pfalzner2015, Hands2019}. The second phase involves the ejection of many of the remaining  planetesimals (rocky and icy), due to close encounters with giant planets during planet migration \citep{Duncan1987, Charnoz2003, Raymond2018}. The final phase is the ejection of the Oort cloud as a star transitions to a white dwarf
\citep{Veras2011, Veras2014, Do2018, Moro-Martin2019}. While icy planetesimals are expected to be ejected across a star's lifetime, rocky planetesimals are thought to be ejected primarily during the gas clearing and planet migration phase, at planetary system ages 100 - 700 Myr \citep{Pfalzner2019}.

The stellar halo is home to the oldest stars in the Milky Way and in the universe \citep{Helmi2008, Frebel2007}. Studying the composition of their ejecta can therefore reveal the formation history of some of the first planets, as well as the enrichment history of the early universe. For instance, the observed ratio of icy to rocky planetesimals can help constrain models of planetary system formation, the chemical composition of rocky planetesimals could reveal unknown details about the gas-clearing and planet migration phases, and the ratios of volatiles in icy planetesimals  could inform the prospects for life in early planetary systems. Compositional information for halo meteors could also help constrain the origins of exo-Oort clouds, as well as chemical diversity among early planetary systems. Furthermore, analysis of halo meteors could test the possibility that carbon-enhanced metal-poor poor stars could create carbon planets \citep{Mashian2016}.

\subsection{Constraints on Baryonic Dark Matter Mass}

Massive compact halo objects (MACHOs) were theorized to partially explain dark matter in the Galactic halo. Constraints on MACHOs from microlensing, wide binaries, and disk kinematics constrain abundances of MACHOs at masses $m \geq 10^{-7} M_{\odot}$
\citep{Alcock1998, Monroy-Rodriguez2014, Brandt2016}. Halo meteors could provide tight constraints on the baryonic dark matter mass fraction for MACHOs at masses $m \leq 10^{-24} M_{\odot}$.

For instance, integrating over the inferred interstellar meteor size distribution \citep{Siraj2019c} implies total mass $m \sim 10^{-5} M_{\odot}$, a few Earth masses, per star of interstellar baryonic material at masses $m \leq 10^{-24} M_{\odot}$. Applying a similar approach to the observed size distribution of halo meteors will allow for tight constraints on baryonic dark matter masss for MACHOs at asteroidal sizes.

\section{Discussion}
\label{sec:discussion}

We analyzed the kinematics of halo meteors, showing their mean heliocentric impact speed to be $\sim 270 \mathrm{\;km\;s^{-1}}$, with a standard deviation of $\sim 90 \mathrm{\;km\;s^{-1}}$. We then explored their abundance and detectability, finding that a worldwide network of all-sky cameras could take spectra and determine orbits for hundreds of $d\geq2$mm halo meteors per year. Finally, we explored the implications of halo meteors on early planetary systems and on constraining baryonic dark matter.

%\pagebreak
%\newpage
\section*{Acknowledgements}
%\vspace{0.1in} 
This work was supported in part by a grant from the Breakthrough Prize Foundation. %\newline \newline

\bsp	% typesetting comment
\label{lastpage}
\end{document}